\documentclass[prl,aps,showpacs, twocolumn]{revtex4}
{
\newfont{\twelvemsb}{msbm10 scaled\magstep1}
\newfont{\eightmsb}{msbm8} \newfont{\sixmsb}{msbm6} \newfam\msbfam
\textfont\msbfam=\twelvemsb \scriptfont\msbfam=\eightmsb
\scriptscriptfont\msbfam=\sixmsb \catcode`\@=11
\def\Bbb{\ifmmode\let\next\Bbb@\else \def\next{\errmessage{Use
\string\Bbb\space only in math mode}}\fi\next}
\def\Bbb@#1{{\Bbb@@{#1}}} \def\Bbb@@#1{\fam\msbfam#1}
\newfont{\twelvegoth}{eufm10 scaled\magstep1}
\newfont{\tengoth}{eufm10} \newfont{\eightgoth}{eufm8}
\newfont{\sixgoth}{eufm6} \newfam\gothfam
\textfont\gothfam=\twelvegoth \scriptfont\gothfam=\eightgoth
\scriptscriptfont\gothfam=\sixgoth \def\frak{\frak@}
\def\frak@#1{{\fam\gothfam{{#1}}}} \def\frak@@#1{\fam\gothfam#1}
\catcode`@=12
\def\CC{{\Bbb C}}
\begin{document}
\title{{\mathversion{bold} {\Large {\bf A three-parametric deformation of
$GL(1/1)$}}}}
\author{Nguyen Anh Ky}
\affiliation{Institute of Physics and Electronics, 10 Dao Tan, Hanoi, Vietnam}
\author{Nguyen Hong Van}
\affiliation{Institute of Physics and Electronics, 10 Dao Tan, Hanoi, Vietnam}
\date{\today}
\begin{abstract}
A three-parametric $R$-matrix satisfying a graded Yang-Baxter equation is introduced.
This $R$-matrix allows us to construct new quantum supergroups which are deformations of
the supergroup $GL(1/1)$ and the universal enveloping algebra $U[gl(1/1)]$. 
 \pacs{1} 
\end{abstract}
\maketitle


Using the $R$-matrix formalism \cite{frt} is one of the approaches
to quantum groups which can be interpreted as a kind of (quantum)
deformations of ordinary (classical) groups or algebras. It has
proved to be a powerful method in investigating quantum groups and
related topics. A physical meaning of this approach is the
so-called (universal) $R$-matrix associated to a quantum group
satisfies the famous Yang-Baxter equation (YBE) representing an
integrability condition of a physical system. The mathematical
advantage of this approach is both the algebraic and co-algebraic
structure of the corresponding quantum group can be expressed in a
few compact (matrix) relations. Quantum groups as symmetry groups of
quantum spaces \cite{frt, manin1} or as deformations of universal
enveloping algebras \cite{drin, jim} can be also derived in an
elegant way in the framework of the $R$-matrix formalism. Combined 
with the supersymmetry idea, the quantum deformations lead to the 
concept of quantum supergroups \cite{manin2, kuresh, svz, chai}. 
In this case, an $R$-matrix becomes graded and satisfies a  
graded YBE. 

By construction, a quantum (super) group depends on one or more,
complex in general, parameters. For about two decades quantum
groups have been investigated in great detail in many aspects.
These investigations were carried out first and mainly on the
one-parametric case and they were extended later to on the
multi-parametric deformations \cite{manin1, reshe}.
Having in principle reacher structures, multi-parametric quantum 
groups are also a subject of interest of a number of authors (see 
\cite{swz, dabrow, pq96, dobrev, pq00, pq01, sl2pq} and references 
therein) and have been applied to considering some physics models (see 
in this context, for example, some recent works, Refs. \cite{sl2pq, kundu, 
jellal, algin}) but in comparison with the one-parametric 
quantum groups, they are considerably less understood (even, in some 
cases they can be proved to be equivalent to one-parametric deformations).
Moreover, most of the multi-parametric deformations considered so
far are two-parametric ones including those of supergroups 
\cite{dabrow, ritten, pq96, dobrev, pq00, pq01} (it is clear that 
two-parametric deformations of supergroups cannot be always reduced 
to one-parametric ones \cite{dabrow, ritten, pq96, pq00}). 
In particular, a two-parametric deformation of the supergroup $GL(1/1)$ 
was considered in \cite{dabrow}. The authors obtained a two-parametric
quantum deformation of $GL(1/1)$ but the corresponding deformation
of the universal enveloping algebra $U[gl(1/1)]$ can be made to
look like an one-parametric deformation by re-scaling its generators 
appropriately. Indeed, starting from the defining relations of the 
deformation of  $U[gl(1/1)]$ given in \cite{dabrow}, 
\begin{eqnarray*}
[K,H]=0,~ [K,\chi_\pm]=0, ~ [H,\chi_\pm]=\pm 2\chi_\pm,
\end{eqnarray*}
\begin{eqnarray*}
\{\chi_+,\chi_-\}_{q/p}=\left(\frac{q}{p}\right)^{H/2}[K]_{qp}~,
\end{eqnarray*}
where
\begin{eqnarray*}
\{\chi_+,\chi_-\}_{q/p}\equiv \left(\frac{q}{p}\right)^{1/2}\chi_+\chi_-
+\left(\frac{q}{p}\right)^{-1/2}\chi_-\chi_+,
\end{eqnarray*}
\begin{eqnarray*}
[K]_{qp}=\frac{(qp)^{K/2}-(qp)^{-K/2}}{(qp)^{1/2}-(qp)^{-1/2}}
\end{eqnarray*}
and making re-scaling $\chi_\pm \rightarrow \chi'_\pm =
\left(\frac{q}{p}\right)^{-1/4}\chi_\pm$, we get 
\begin{eqnarray*}
[K,H]=0, [K,\chi'_\pm]=0, [H,\chi'_\pm]=\pm 2\chi'_\pm,
\{\chi'_+,\chi'_-\}=[K]_{qp}.
\end{eqnarray*}
The latter relations are (conventional) defining relations of an one-parametric deformation
of $U[gl(1/1)]$ with parameter $\sqrt{qp}$. In the present paper
we suggest an $R$-matrix allowing us to construct  a
three-parametric deformation of $GL(1/1)$. This suggestion,
however, is two-fold, as it leads us to a true two-parametric
deformation of $U[gl(1/1)]$.\\


  Let us start with the operator
\begin{eqnarray}
R&=&q(e^1_1\otimes e^1_1) + r(e^1_1\otimes e^2_2)+ s(e^2_2\otimes e^1_1)+
\lambda(e^1_2\otimes e^2_1)\nonumber\\
&& + p(e^2_2\otimes e^2_2),
\label{r}
\end{eqnarray}
where $p$, $q$, $r$, $s$ and $\lambda$ are complex deformation parameters ($p, q, r, s, \lambda \in \CC$), 
while $e^i_j$, $i,j=1,2$, are Weyl generators of $GL(1 | 1)$ with
a $Z_2$-grading given as follows:
\begin{eqnarray}
[e^i_j]=[i]+[j]~ (\mbox{mod 2}),~~ [i]=\delta_{i2}.
\end{eqnarray}
We call the latter operator an $R$-matrix although it has a
(finite) matrix form only in a finite-dimensional representation.
In the fundamental representation $e^i_j$ are super-Weyl matrices,
$(e^i_j)^h_k=\delta^i_k\delta^h_j$, and $R$ is a $4\times 4$
matrix. Three of the five parameters, say, $p$, $q$ and $r$, can
be chosen to be independent, while the remaining parameters, $s$
and $\lambda$, are subject to the constraints
\begin{eqnarray*}
rs=pq, ~~ \lambda=q-p.
\end{eqnarray*}
By this choice of the parameters, the $R$-matrix (1)
satisfies the graded YBE
\begin{eqnarray}
R_{12}R_{13}R_{23}=R_{23}R_{13}R_{12},
\end{eqnarray}
with
\begin{eqnarray}
R_{12}&=&R\otimes I\equiv R\otimes e^i_i,~~ i=1,2,\nonumber \\
R_{13}&=&q(e^1_1\otimes e^i_i \otimes e^1_1) + r(e^1_1\otimes e^i_i \otimes e^2_2)+
s(e^2_2\otimes e^i_i \otimes e^1_1)\nonumber \\
&&+(-1)^{[i]}\lambda(e^1_2\otimes e^i_i \otimes e^2_1)
+ p(e^2_2\otimes e^i_i \otimes e^2_2),\nonumber \\
R_{23}&=&I\otimes R\equiv e^i_i\otimes R,
\end{eqnarray}
where repeated indices are summation indices, $I$ is the identity
operator and the $Z_2$-grading is given in (2).\\

Now suppose the operator subject
\begin{eqnarray}
T= a~e^1_1+\beta ~e^1_2+\gamma ~e^2_1+d~e^2_2\equiv t^j_ie^i_j,
\label{t}
\end{eqnarray}
which in the fundamental representation is a $2\times 2$ matrix,
\begin{eqnarray}
T=\left (\begin{tabular}{ccccc}
$a$&&&&$\beta$\\ $\gamma$&&&&$d$
\end{tabular}\right ),
\label{tfrep}
\end{eqnarray}
obeys the so-called $RTT$ equation
\begin{eqnarray}
RT_1T_2=T_2T_1R,
\label{rtt}
\end{eqnarray}
where
\begin{eqnarray}
T_1&=& T\otimes I\equiv (a e^1_1+\beta e^1_2+\gamma e^2_1+d e^2_2)\otimes e^i_j, 
\nonumber \\
T_2&=&I\otimes T\nonumber \\
&\equiv & e^i_j\otimes [ a e^1_1+(-1)^{[i]}\beta e^1_2+(-1)^{[i]}
\gamma e^2_1+d e^2_2].
\end{eqnarray}
The Eq. (\ref{rtt}) leads to the
commutation relations between the elements of $T$ :
\begin{eqnarray}
a\beta &=&\frac{r}{p}\beta a,~  a\gamma =\frac{q}{r}\gamma a,~ ad=da+
\frac{\lambda}{r}\gamma \beta,~ \beta^2=0=\gamma^2,
\nonumber 
\\
\beta\gamma &=&-\frac{s}{r}\gamma\beta\equiv -\frac{pq}{r^2}\gamma\beta ,
~ \beta d=\frac{p}{r}d\beta , ~ \gamma d=\frac{r}{q}d\gamma .
\label{tcr}
\end{eqnarray}
Let us denote $G$ a set of all
operators (\ref{t}) satisfying (\ref{rtt}) and let $T$ and  $T'$ be two independent
copies of (\ref{t}) in the sense that all elements $t^i_j$ of $T$
commute with all those of $T'$. The fact that the multiplication
$T.T'$  preserves the relation (\ref{rtt}), that is, the relations (\ref{tcr}), reflects the
group nature of $G$. Next, since the quantity
\begin{eqnarray}
D(T) &\equiv & (a-\beta d^{-1}\gamma )d^{-1}=d^{-1}(a-\beta d^{-1}
\gamma)\nonumber \\
&=&a(d-\gamma a^{-1}\beta)^{-1}
\end{eqnarray}
commutes with $T$ and has the "multiplicative"
property $D(T.T')=D(T).D(T')$ it can be identified with a
representation of a {\bf quantum superdeterminant}. Thus we can
take $G$ with $D(T)\neq 0$, $\forall T\in G$, as a
three-parametric deformation, denoted by $GL_{p,q,r}(1/1)$, of a
representation of $GL(1/1)$. In the fundamental representation,
$D(T)$ is a quantum superdeterminant of matrix $T$ in (\ref{tfrep}) 
and the corresponding $GL_{p,q,r}(1/1)$ is a three-parametric 
deformation of $GL(1/1)$. When we set $D(T)=1$ we get a 
three-parametric deformation of $SL(1/1)$. We note that the form  of 
$D(T)$ is the same as in \cite{dabrow}, that is, it remains non-deformed 
and belongs to the center of $GL_{p,q,r}(1/1)$. The
Hopf structure is straightforward and given by the following maps:\\

- the co-product: 
\begin{eqnarray}
\Delta (T) = T \dot{\otimes} T,
\end{eqnarray}

- the antipode: 
\begin{eqnarray}
S(T).T=I,
\end{eqnarray}

- the counit:  
\begin{eqnarray}
\varepsilon (T) = I.
\end{eqnarray}
In components they read
\begin{eqnarray}
\Delta (t^i_j)=t^k_j\otimes t^i_k,
\end{eqnarray}
\begin{eqnarray}
S(t^j_ie^i_j) &=&S(t^j_i)e^i_j\nonumber \\
&=& a^{-1}(1+\beta d^{-1}\gamma a^{-1})e^1_1-(a^{-1}\beta d^{-1})e^1_2
\nonumber \\
&&-(d^{-1}\gamma a^{-1})e^2_1+ d^{-1}(1-\beta a^{-1}\gamma d^{-1})e^2_2,~~~~~
\end{eqnarray}
\begin{eqnarray}
\varepsilon(t^i_j)=\delta^i_j.
\end{eqnarray}
A quantum superplane with symmetry (authomorphism) group 
$GL_{p,q,r}(1/1)$ is given by the coordinates
\begin{eqnarray}
\left (\begin{tabular}{c}
$x$\\ $\theta$
\end{tabular}\right )  ~~~\mbox{or} ~~~ \left (\begin{tabular}{c}
$\eta$\\$y$
\end{tabular}\right )
\end{eqnarray}
subject to the commutation relations
\begin{eqnarray}
x\theta= {q\over r}\theta x\equiv {s\over p}\theta x,~ \theta^2=0 
~~~ {\rm or} ~~~
\eta^2=0,~ \eta y={p\over r}y\eta, 
\end{eqnarray}
respectively. Note that these quantum superplanes (which are 
"two-dimensional") are still two-parametric (of course, we cannot 
make relations between two coordinates to depend on
more than two parameters). Finally, in order to complete 
our program we must conctruct a deformation, denoted below as 
$U_{p,q,r}[gl(1/1)]$, of the universal enveloping algebra
$U[gl(1/1)]$ corresponding to the $R$-matrix (\ref{r}).\\


First, we introduce two auxilary operators
\begin{eqnarray}
L^+&=&H_1^+e^1_1+H_2^+e^2_2+\lambda X^+e^1_2, \nonumber \\
L^-&=&H^-_1e^1_1+H^-_2e^2_2+\lambda X^-e^2_1,
\end{eqnarray}
with $H_i^{\pm}$ and $X^{\pm}$ belonging to $U_{p,q,r}[gl(1/1)]$
to be constructed. Then, demanding
\begin{eqnarray}
L^{\pm}_1&=&L^{\pm}\otimes e^i_i,
\nonumber\\
L^+_2&=&e^i_i\otimes [H_1^+e^1_1+H_2^+e^2_2+(-1)^{[i]}\lambda X^+e^1_2],
\nonumber \\
L^-_2&=&e^i_i\otimes [H^-_1e^1_1+H^-_2e^2_2+(-1)^{[i]}\lambda X^-e^2_1]
\end{eqnarray}
to obey the equations
\begin{eqnarray}
RL^{\epsilon_1}_1L^{\epsilon_2}_2=L^{\epsilon_2}_2L^{\epsilon_1}_1R,
\end{eqnarray}
where $(\epsilon_1,\epsilon_2)=(+,+), (-,-),(+,-)$, we get the
following commutation relations between $H_i^{\pm}$ and $X^{\pm}$:
\begin{eqnarray}
H_i^{\epsilon_1}H_j^{\epsilon_2}=H_j^{\epsilon_2}H_i^{\epsilon_1},
\nonumber 
\end{eqnarray}
\begin{eqnarray}
pH_i^+X^+=rX^+H_i^+, ~~qH^-_iX^+=rX^+H^-_i,
\nonumber 
\end{eqnarray}
\begin{eqnarray}
rH_i^+X^-=pX^+H_i^+, ~~rH^-_iX^-=qX^-H^-_i,
\nonumber 
\end{eqnarray}
\begin{eqnarray}
rX^+X^-+sX^-X^+=\lambda^{-1}(H^-_2H_1^+-H_2^+H^-_1),
\label{ugl}
\end{eqnarray}
which are taken to be the defining relations of $U_{p,q,r}[gl(1/1)]$.
Its Hopf structure is given by
\begin{eqnarray}
\Delta(L^{\pm})&=&L^{\pm}\dot{\otimes}L^{\pm},
\\
S(L^{\pm})&=&(L^{\pm})^{-1},
\\
\varepsilon(L^{\pm})&=&I,
\end{eqnarray}
or equivalently (no summation on $i=1,2$),
\begin{eqnarray}
\Delta(H_i^{\pm})&=&H_i^{\pm}\otimes H_i^{\pm},
\nonumber 
\\
\Delta(X^+)&=&H_1^+\otimes X^+ +X^+\otimes H_2^+,
\nonumber 
\\
\Delta(X^-)&=&H_2^-\otimes X^- +X^-\otimes H_1^-, 
\end{eqnarray}
\begin{eqnarray}
S(H_i^{\pm})&=&(H_i^{\pm})^{-1}, 
\nonumber 
\\
S(X^+)&=&-(H_1^+)^{-1}X^+(H_2^+)^{-1},
\nonumber 
\\
S(X^-)&=&-(H_2^-)^{-1}X^-(H_1^-)^{-1},
\end{eqnarray}
\begin{eqnarray}
\varepsilon(H_i^{\pm})=1, ~~ \varepsilon(X^{\pm})=0.
\end{eqnarray}
At first sight $U_{p,q,r}[gl(1/1)]$ given in (\ref{ugl}) is a
three-parametric quantum supergroup depending on three parameters
$p$, $q$ and $r$ (or $s$). However, making the substitution
\begin{eqnarray}
H_1^+=\left ( {r\over p}\right)^{E_{11}}, ~~ H_2^+
=\left ({p\over r}\right)^{E_{22}}, 
\nonumber\\
H^-_1=\left ( {r\over q}\right)^{E_{11}}, ~~ H^-_2= \left ({q\over
r}\right)^{E_{22}}, \nonumber\\
E_{12}=X^+r^{E_{22}}, ~~ E_{21}=X^-s^{E_{11}},
\label{reali}
\end{eqnarray}
we obtain a two-parametric deformation of $U[gl(1/1)]$, namely,
\begin{eqnarray}
[E_{ii},E_{jj}] &=& 0,
\nonumber 
\end{eqnarray}
\begin{eqnarray}
[ E_{ii},E_{j,j\pm 1}]&=& (\delta_{ij}-\delta_{i,j\pm 1}) E_{j,j\pm 1},
\nonumber 
\end{eqnarray}
\begin{eqnarray}
\{E_{12},E_{21}\}&=&[K]_{q,p}~,
\end{eqnarray}
where
$1\le i,j,j\pm1\le 2$ and
\begin{eqnarray}
[K]_{q,p}=\frac{q^K-p^K}{q-p}, ~~ K=E_{11}+E_{22}.
\end{eqnarray}
The latter deformation is a true two-parametric deformation of $U[gl(1/1)]$ as
it cannot be made to become one-parametric by rescaling its
generators. Of course, (\ref{reali}) is not the only realization of the
generators of $U_{p,q,r}[gl(1/1)]$ in terms of the deformed Weyl
generators $E_{ij}$.\\


We have suggested in the present paper an $R$-matrix which satisfies a 
three-parametric graded YBE (and modified Hecke conditions which are not 
exposed here). Using this $R$-matrix we obtained new deformations of 
$GL(1/1)$ and $U[gl(1/1)]$. For conclusion, let us emphasize that the 
deformation $GL_{p,q,r}(1/1)$ of $GL(1/1)$ obtained is a three-parametric
quantum group despite the fact that the corresponding deformation
$U_{p,q,r}[gl(1/1)]$ is equivalent to a two-parametric deformation
of $U[gl(1/1)]$. On the other side, however, the introduction of
the latter solves a small problem of \cite{dabrow} mentioned
above.\\

\begin{acknowledgements}
This work was supported in part by the National Research
Program for Natural Sciences of Vietnam under Grant No  410804.
\end{acknowledgements}


\begin{thebibliography}{}
\bibitem{frt} L. Faddeev, N. Reshetikhin and L. Takhtajan,
Algebra and Analys {\bf 1}, 178  (1987).
\bibitem{manin1} Yu. Manin, {\it Quantum groups and  non-commutative geometry},
Centre des Recherchers  Math\'ematiques, Montr\'eal, 1988.
\bibitem{drin} V. Drinfel'd, "Quantum groups", J. Sov. Math., {\bf 41},
898 (1988); {\it Zap. Nauch. Semin.} {\bf 155}, 18 (1986); also in
{\it  Proceedings of the International Congress of  Mathematicians,
Berkeley 1986}, vol {\bf 1}, The American Mathematical Society,
Providence, RI, 1987,  pp. 798 - 820.
\bibitem{jim} M. Jimbo, Lett. Math. Phys. {\bf 10},
63 (1985); {\it ibit} {\bf 11}, 247 (1986).
\bibitem{manin2} Yu. Manin, Commun. Math. Phys. {\bf 123}, 169 (1989).
\bibitem{kuresh}P. Kulish and N. Reshetikhin, Lett. Math. Phys {\bf 18},  143 (1989).
\bibitem{svz} J. Schmidke, S. Volos and B. Zumino, Z. Phys. C {\bf 48}, 249 (1990).
\bibitem{chai} M. Chaichian and P. Kulish, Phys. Lett. {\bf 234B}, 72 (1990).
\bibitem{reshe} N. Reshetikhin, Lett. Math. Phys. {\bf 20}, 331 (1990).
\bibitem{swz} A. Schirrmacher, J. Wess and B. Zumino, Z. Phys. C
{\bf 49}, 317 (1991).
\bibitem{dabrow} L. Dabrowski and L.-y. Wang, Phys. Lett. {\bf 266B}, 51 (1991).
\bibitem{ritten} H. Hinrichsen and V. Rittenberg, Phys. Lett. {\bf 275B}, 350 (1992); hep-th/9110074.
\bibitem{pq96} Nguyen Anh Ky, J. Phys. A {\bf 29}, 1541 (1996)
or math.QA/9909067.
\bibitem{dobrev} V. Dobrev and E. Tahri, Int. J. Mod. Phys. A {\bf 13}, 4339 (1998).
\bibitem{pq00} Nguyen Anh Ky, J. Math. Phys. {\bf 41},
6487 (2000); math.QA/0005122.
\bibitem{pq01} Nguyen Anh Ky, J. Phys. A
{\bf 34}, 7881 (2001);  math.QA/0104105.
\bibitem{sl2pq} Nguyen Anh Ky and Nguyen Thi Hong Van,
"A two-parametric deformation of U[sl(2)], its representations and
complex "spin"", math.QA/0506539.
\bibitem{kundu} A. Kundu, Phys. Rev. Lett. {\bf 82}, 3936 (1999).
\bibitem{jellal} A Jellal, Mod. Phys. Lett. A {\bf 17}, 701 (2002).
\bibitem{algin} A. Algin and B Deriven, J. Phys. A {\bf 38}, 5945 (2005).
\end{thebibliography}
\end{document}